\documentclass[aps, prev, twocolumn, preprintnumbers, floatfix, nofootinbib]{revtex4-1}

\usepackage{graphicx}
\usepackage[english]{babel}
\usepackage[utf8]{inputenc}
\usepackage{bm}
\usepackage{times}
\usepackage{slashed}
\usepackage{mathtools}
\usepackage{epsfig}
\usepackage{dcolumn}
\usepackage{textcomp}
\usepackage{color}
\usepackage{verbatim}
\usepackage{listings}
\usepackage{xfrac}

\def\hmath$#1${\texorpdfstring{{\rmfamily\textit{#1}}}{#1}}

\usepackage{xcolor}
\definecolor{oxfordblue}{RGB}{0,33,70}
\usepackage{hyperref}
\hypersetup{
     urlcolor=oxfordblue,
     linktocpage=true,
     setpagesize=true,
     urlcolor=oxfordblue,
     citecolor=oxfordblue,
     linkcolor=oxfordblue, 
     menucolor=cyan,
     colorlinks=true, 
     filecolor=magenta,
     citebordercolor=blue,
     pdftitle={Search for a Doubly Charged Scalar at the LHC and FCC-hh},
     pdfsubject={Latex},
     pdfkeywords={BSM},
     unicode=true,
}

\usepackage{soul}
\usepackage{subfigure}
\usepackage[capitalize, nameinlink]{cleveref}

\usepackage{amsfonts}
\usepackage{threeparttable}

\begin{document} 

\title{Search for a Doubly Charged Scalar at the LHC and FCC-hh}

\author{Letícia Guedes$^{1,2,3}$}
\author{Gabriela Hoff$^{1}$}
\author{Farinaldo S. Queiroz$^{1,2,3,4}$}
\email{farinaldo.queiroz@ufrn.br}
\author{Y.M. Oviedo-Torres $^{2,5}$}
\author{Y. Villamizar$^{6}$}

\affiliation{$^1$International Institute of Physics, Universidade Federal do Rio Grande do Norte, Campus Universitário, Lagoa Nova, Natal-RN 59078-970, Brazil}
\affiliation{$^2$Millennium Institute for Subatomic Physics at High-Energy Frontier (SAPHIR), Fernandez Concha 700, Santiago, Chile.}
\affiliation{$^3$Departamento de F\'{\i}sica, Universidade Federal do Rio Grande do Norte, 59078-970, Natal, RN, Brasil}
\affiliation{$^4$Departamento de F\'isica, Facultad de Ciencias, Universidad de La Serena, Avenida Cisternas 1200, La Serena, Chile}
\affiliation{$^5$Center for Theoretical and Experimental Particle Physics - CTEPP,
Facultad de Ciencias Exactas, Universidad Andrés Bello, Fernández Concha 700, Santiago, Chile}
\affiliation{$^6$Laboratoire de Physique Théorique et Hautes Énergies (LPTHE), UMR 7589, Sorbonne Université et CNRS, 4 place Jussieu, 75252 Paris Cedex 05, France}


\begin{abstract}
Doubly charged scalars frequently emerge in many well-motivated extensions of the Standard Model, particularly in frameworks that aim to explain the origin of neutrino masses. Their distinct electric charge and clean leptonic signatures make them especially compelling from the standpoint of experimental searches. In this work, we explore the sensitivity of the LHC full run II, including photon-photon fusion, and Future Circular Collider in its hadron-hadron configuration (FCC-hh) to such states, assuming they decay promptly and exclusively into charged leptons—either conserving or violating lepton flavor. We find that the FCC-hh, operating at 100 TeV, is uniquely positioned to probe doubly charged scalars with masses up to 7 TeV and possibly establish the mechanism behind neutrino masses. 
\end{abstract}

\maketitle
\flushbottom

\section{Introduction \label{SecI_introd}}

Doubly charged scalar fields appear in a wide range of extensions of the Standard Model. They are especially well known from the type‑II seesaw framework \cite{Mohapatra:1974hk, Senjanovic:1975rk}, where an additional SU(2) scalar triplet is introduced to generate neutrino masses. This triplet necessarily includes a doubly charged component \cite{Muhlleitner:2003me, Akeroyd:2005gt, Hektor:2007uu, FileviezPerez:2008wbg, Chaudhuri:2013xoa, Lindner:2016bgg, Primulando:2019evb}, giving rise to characteristic collider signatures. Several models with extended scalar sectors also predict such states, including the Left--Right Model \cite{Pati:1974yy, Dutta:2014dba, Dev:2016vle, Borah:2016hqn, Dev:2018kpa}, the 3-3-1 Model \cite{CiezaMontalvo:2006zt, Alves:2011kc, Alves:2012yp, Machado:2018sfh, CarcamoHernandez:2019vih, Ferreira:2019qpf}, little‑Higgs models \cite{ArkaniHamed:2002qx, Hektor:2007uu}, the Georgi--Machacek model \cite{Georgi:1985nv, Sun:2017mue, Chiang:2018cgb}, and other scalar triplet extensions \cite{Zee:1985id, Gunion:1989ci, Nebot:2007bc, Akeroyd:2012ms, deMedeirosVarzielas:2017glw, Ghosh:2018jpa, Camargo:2018uzw, Chala:2018opy, Chabab:2018ert, Chakraborty:2019uxk}. Therefore, doubly charged scalars constitute a recurring feature in beyond-Standard-Model physics.

This has motivated several collider searches at hadron colliders \cite{Huitu:1996su, Akeroyd:2005gt, Han:2007bk, Chao:2008mq, Akeroyd:2010ip, Chiang:2012dk, delAguila:2013mia, Kanemura:2013vxa, King:2014uha, kang:2014jia, Kanemura:2014ipa, Bambhaniya:2015wna, Mitra:2016wpr, Du:2018eaw}, followed by dedicated searches by the CMS and ATLAS collaborations \cite{CMS:2012kua, ATLAS:2012hi, Chatrchyan:2012ya, ATLAS:2014kca, ATLAS:2016pbt, CMS:2016cpz, CMS:2017pet, Aaboud:2017qph}. The main signature consists of highly energetic same-sign charged leptons produced from the decay of the doubly charged scalar. The large momentum transfer and high invariant mass of these leptons provide excellent discrimination against Standard Model (SM) backgrounds, making the signal remarkably clean. Although no excess of events has been observed to date, lower mass limits have been imposed, ranging from 600 to 800 GeV, depending on the interaction structure and decay channels.

\begin{figure}
    \centering
    \includegraphics[width=0.4\textwidth]{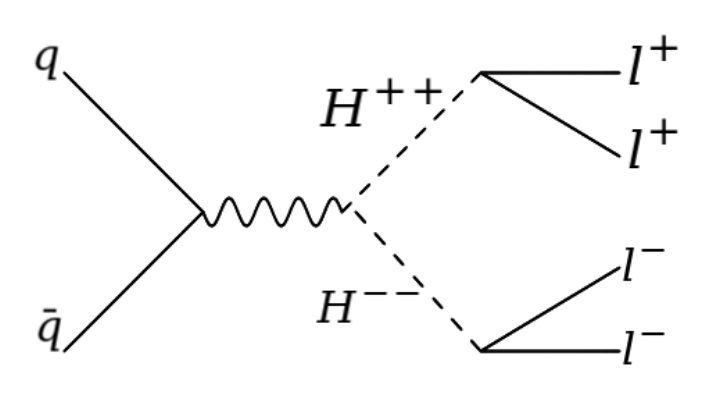}
    \caption{\label{fig:diagram1} Feynman diagram of the resonant pair production process of the doubly charged scalar decaying into charged lepton pairs.}
\end{figure}

It is often assumed that doubly charged scalars couple only to left-handed fermions, but this is not always the case in Left--Right models. Regardless, it is reasonable to assume that the dominant decays occur into charged leptons. In type‑II seesaw constructions, this is ensured when the vacuum expectation value (vev) of the scalar triplet is sufficiently small \cite{Perez:2008ha}. For vevs of order MeV, the decay $H^{\pm\pm}\rightarrow W^{\pm}W^{\pm}$ becomes relevant, though it leads to weaker bounds \cite{ATLAS:2018ceg}. Doubly charged scalars can be produced in pairs or in association with singly charged scalars, but pair production generally yields stronger limits due to its larger signal-to-background ratio. Thus, we focus on the resonant pair-production channel, as in the CMS and ATLAS analyses \cite{CMS:2012kua, ATLAS:2012hi, Chatrchyan:2012ya, ATLAS:2014kca, ATLAS:2016pbt, CMS:2016cpz, CMS:2017pet, Aaboud:2017qph}. The relevant Feynman diagram is shown in \Cref{fig:diagram1}.

A key assumption is the narrow-width approximation, which holds when the decay width is small compared to the particle mass ($\Gamma \ll M$), the decay products are much lighter than the parent particle, and interference with non-resonant processes is negligible \cite{delaMadrid:2017oeo, Berdine:2007uv}. These conditions are satisfied for the resonant production of a doubly charged scalar shown in \Cref{fig:diagram1}. While large-width effects introduce mild changes to the mass bounds \cite{Crivellin:2018ahj}, for heavy masses ($m_{H^{\pm\pm}} \gtrsim 1.5$ TeV), photon--photon fusion—enhanced by the scalar's trilinear and quartic couplings to photons—can contribute significantly \cite{Babu:2016rcr, Crivellin:2018ahj}.

The latest mass bounds were obtained using an integrated luminosity of $\mathcal{L}=36.1$ fb$^{-1}$ \cite{ATLAS:2017xqs}. Our first objective is to update these limits to $\mathcal{L}=139$ fb$^{-1}$ for various decay channels. For future colliders, the FCC-hh operating at 100 TeV promises an exceptional reach, pushing direct searches to the next energy frontier \cite{FCC:2025lpp, FCC:2025uan, FCC:2025jtd}. The recent update of the European Strategy for Particle Physics clearly identifies FCC-hh as the most promising machine in terms of physics potential \cite{deBlas:2025gyz}. Motivated by this, we also forecast its reach for doubly charged scalars.

This paper is organized as follows: \Cref{SecII_Model} describes the model; \Cref{SecIII_Ana} presents the LHC bounds; \Cref{SecIV_FCC} contains the FCC-hh projections; and \Cref{SecV_Con} summarizes our conclusions.

\section{Model \label{SecII_Model}}


We consider a type-II seesaw scenario in which a complex scalar triplet extends the Standard Model $\Delta \sim (\mathbf{3},\, Y=2)$ under $SU(2)_L\times U(1)_Y$. The Lagrangian describing the relevant interactions is the following,

\begin{equation}
 \mathcal{L} \supset \text{Tr}[(D_\mu \Delta)^\dagger (D^\mu \Delta)] + y_{ab}\, L^T_a C\, i\sigma_2\, \Delta\, L_b + \text{h.c.},
 \label{Eq:lagrangian1}
\end{equation}

where $L_a=(\nu_a,\ell_a)^T$ denotes the left-handed lepton doublet of flavor $a$, $C$ is the charge-conjugation matrix, and $\sigma_i$ are the Pauli matrices. The triplet field can be written as,

\begin{eqnarray}
\Delta &=& \frac{\sigma^i}{\sqrt{2}} \Delta_i =
\begin{pmatrix}
H^+/\sqrt{2} & H^{++} \\
H^0 & -H^+/\sqrt{2}
\end{pmatrix}.
\end{eqnarray}

The covariant derivative acting on $\Delta$ is

\begin{equation}
D_\mu \Delta = \partial_\mu \Delta + i\frac{g}{2}[\sigma^a W^a_\mu, \Delta] + ig^\prime B_\mu \Delta,
\end{equation}
with $W_\mu^a$ and $B_\mu$ the $SU(2)_L$ and $U(1)_Y$ gauge fields and $g$ and $g^\prime$ their corresponding gauge couplings.

Spontaneous symmetry breaking occurs when the neutral component ($H^0$) acquires a vacuum expectation value (vev) $v_t$, generating neutrino masses $(m_\nu)_{ab}= \sqrt{2}\, v_t\, y_{ab}$, as in the standard type-II seesaw mechanism \cite{BhupalDev:2013xol}. Masses of order $0.1$ eV can be achieved by suppressing either $v_t$ or $y_{ab}$. The triplet vev also contributes to the electroweak gauge boson masses, $M_Z^2=\sfrac{g^2(v_h^2+4v_t^2)}{(4\cos^2\theta_W)}$ and $M_W^2= \sfrac{g^2(v_h^2+2v_t^2)}{4}$ \cite{Ghosh:2017pxl}. Precision measurements of the $\rho$ parameter, $\rho=1.00004$ \cite{ParticleDataGroup:2024cfk}, imply $v_t \lesssim \mathcal{O}(1$--$5)\,$GeV.

The gauge interactions of $H^{\pm\pm}$ arise from the kinetic term in \Cref{Eq:lagrangian1} and are controlled by $v_t$. For $v_t \lesssim 10^{-4}$~GeV, the triplet scalars decay predominantly into leptons, whereas for larger $v_t$, the bosonic decay $H^{\pm\pm}\to W^\pm W^\pm$ may become relevant \cite{Ghosh:2017pxl}. Since the bosonic regime introduces additional model dependence and yields weaker bounds in the analyses we recast, we focus on the leptonic-dominated regime, in which the partial width is \cite{Crivellin:2018ahj}
\begin{equation}
\Gamma(H^{\pm\pm} \rightarrow \ell_a^\pm \ell_b^\pm) = \frac{k\, |y_{ab}|^2}{16\pi} m_{H^{\pm\pm}},
\end{equation}
with $k=1$ for $a\neq b$ and $k=2$ for $a=b$. Following the ATLAS strategy in, we neglect $\tau$ final states, but we allow lepton-flavor violation through off-diagonal Yukawa couplings.

Thus, we consider the benchmark decays $H^{\pm\pm}\to e^\pm e^\pm$, $e^\pm\mu^\pm$, and $\mu^\pm\mu^\pm$. We restrict to prompt decays (hence the narrow-width approximation); displaced decays may occur for $|y_{ab}|\ll 10^{-5}$ \cite{Han:2007bk, Dev:2018kpa, Antusch:2018svb} and are omitted here.

For collider purposes, we adopt a simplified setup in which $m_{H^{\pm\pm}}$ is treated as an independent parameter. We assume the remaining triplet-like states are sufficiently close in mass so that cascade decays such as $H^{\pm\pm}\to H^\pm W^{\pm(*)}$ are kinematically suppressed. Under these assumptions, the dominant production mechanism at hadron colliders is Drell-Yan pair production via $\gamma^*/Z^*$, $pp\to H^{++}H^{--}$.

\section{Analysis \label{SecIII_Ana}}

We implemented the type-II seesaw model described in \Cref{SecII_Model} in \texttt{FeynRules} \cite{Alloul:2013bka} and exported the corresponding UFO model files \cite{Degrande:2011ua}. Event generation was performed at leading order with \texttt{MadGraph5} \cite{Alwall:2011uj} for $pp\to H^{++}H^{--}$ at $\sqrt{s}=13$~TeV.  Throughout, we focus on the resonant pair-production channel followed by prompt leptonic decays,
\begin{equation}
pp\to H^{++}H^{--}\to \ell^+\ell^+\,\ell^-\ell^-,
\end{equation}
with $\ell=e,\mu$, and we consider both flavor-conserving and flavor-violating final states according to the benchmark branching fractions stated below.

Experimental searches for doubly charged scalars are interpreted assuming fixed branching ratios into charged leptons, since the reconstruction efficiencies and background contamination depend sensitively on the lepton flavor composition. Charge misidentification is negligible for muons due to the combined performance of the inner tracking detector and the muon spectrometer. In contrast, for electrons, the dominant source of charge misidentification arises from hard bremsstrahlung followed by photon conversion, which can significantly degrade the charge assignment at high transverse momentum \cite{ATLAS:2017xqs}. As a consequence, electron charge misidentification introduces a systematic uncertainty of order $10\%$ for $p_T \gtrsim 100$~GeV \cite{ATLAS:2014kca}, which is precisely the kinematic regime relevant for heavy doubly charged scalar searches. This effect leads to systematically weaker bounds in dielectron final states compared to dimuon channels.

\begin{figure}[!h]
    \includegraphics[width=0.48\textwidth]{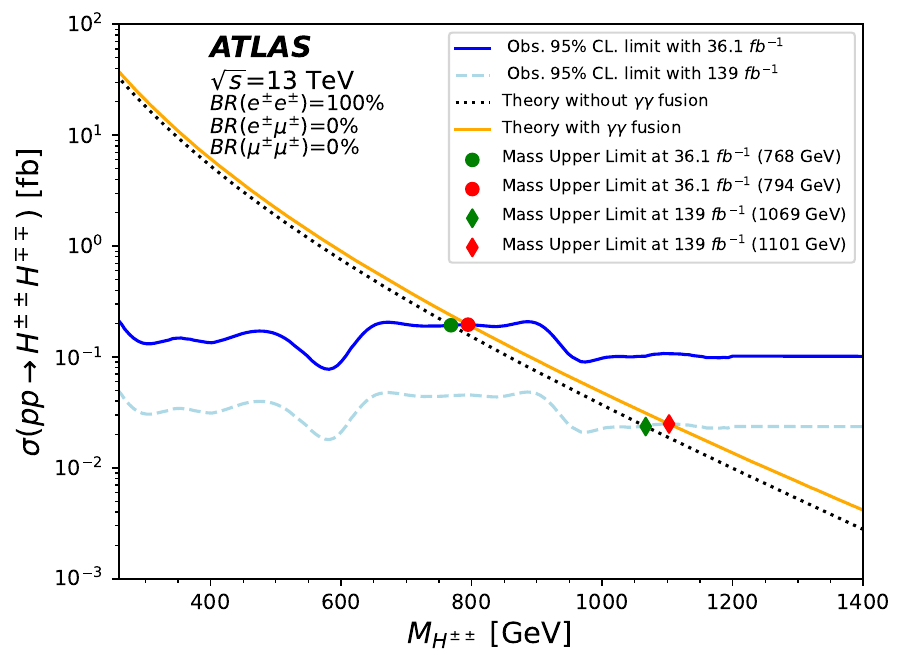}
    \caption{\label{fig:case_100_0_0} Production cross section as a function of the doubly charged scalar mass assuming $\text{BR}(H^{--}\rightarrow ee)=100\%$, $\text{BR}(H^{--}\rightarrow \mu\mu)=0\%$, and $\text{BR}(H^{--}\rightarrow \mu e)=0\%$. We overlay the current and projected ATLAS limits at 13 TeV center-of-mass energy with $36.1\,\text{fb}^{-1}$ and $139\,\text{fb}^{-1}$ of integrated luminosity.}
\end{figure}

\begin{figure}[!h]
    \includegraphics[width=0.48\textwidth]{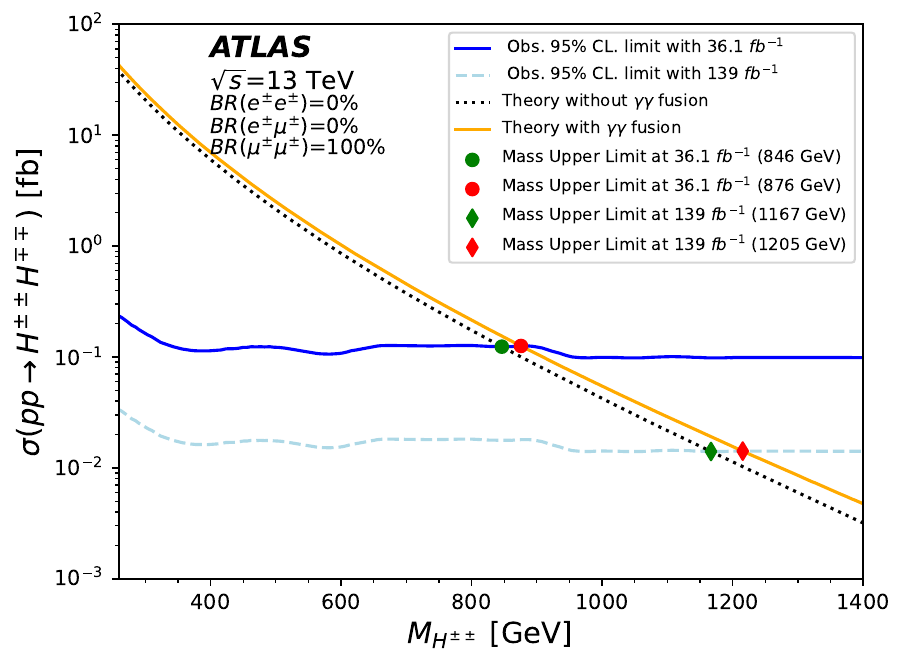}
    \caption{ \label{fig:case_0_100_0} Production cross section as a function of the doubly charged scalar mass assuming $\text{BR}(H^{--}\rightarrow ee)=0\%$, $\text{BR}(H^{--}\rightarrow \mu\mu)=100\%$, and $\text{BR}(H^{--}\rightarrow \mu e)=0\%$.}
\end{figure}

\begin{figure}[!h]
    \includegraphics[width=0.48\textwidth]{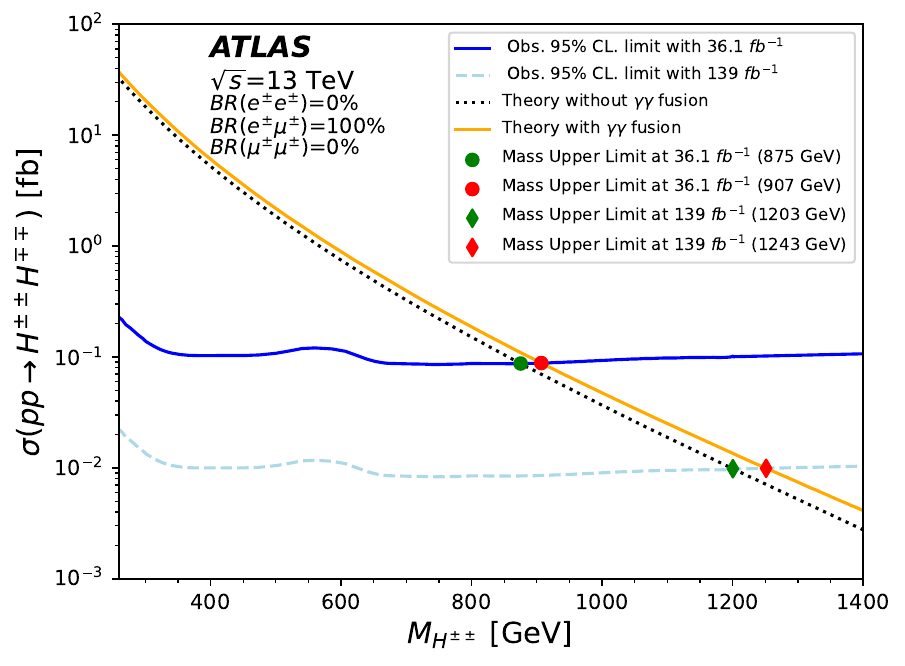}
    \caption{\label{fig:case_0_0_100} Production cross section as a function of the doubly charged scalar mass assuming $\text{BR}(H^{--}\rightarrow ee)=0\%$, $\text{BR}(H^{--}\rightarrow \mu\mu)=0\%$, and $\text{BR}(H^{--}\rightarrow \mu e)=100\%$.}
\end{figure}

\begin{figure}[!h]
    \includegraphics[width=0.48\textwidth]{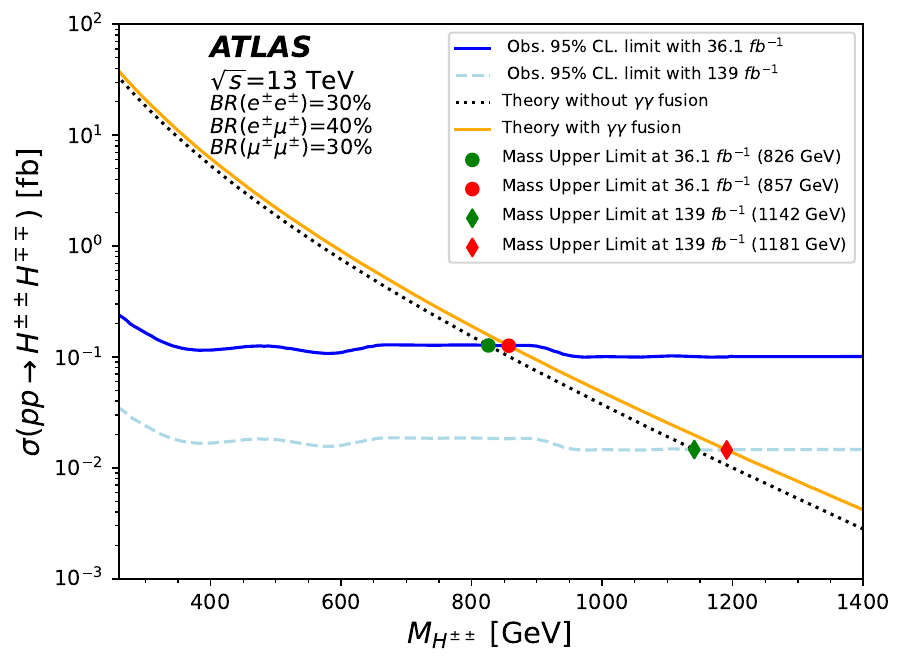}
    \caption{\label{fig:case_30_40_30} Production cross section as a function of the doubly charged scalar mass assuming $\text{BR}(H^{--}\rightarrow ee)=30\%$, $\text{BR}(H^{--}\rightarrow \mu\mu)=40\%$, and $\text{BR}(H^{--}\rightarrow \mu e)=30\%$.}
\end{figure}

\paragraph{Validation against ATLAS.}
To validate our implementation, we computed the inclusive production cross section $\sigma(pp\to H^{++}H^{--})$ as a function of the doubly charged scalar mass and compared it with the 95\% confidence level upper limits reported by the ATLAS collaboration for $\sqrt{s}=13$~TeV and an integrated luminosity of $\mathcal{L}=36.1~\mathrm{fb}^{-1}$ \cite{ATLAS:2017xqs}. We considered the same benchmark decay hypotheses adopted in the experimental analysis,
\begin{itemize}
    \item[(i)] $\text{BR}(H^{\pm\pm} \rightarrow e^{\pm}e^{\pm})=100\%$ in \Cref{fig:case_100_0_0};
    \item[(ii)] $\text{BR}(H^{\pm\pm}\rightarrow \mu^{\pm}\mu^{\pm})=100\%$ in \Cref{fig:case_0_100_0};
    \item[(iii)] $\text{BR}(H^{\pm\pm}\rightarrow e^{\pm}\mu^{\pm})=100\%$ in \Cref{fig:case_0_0_100}; 
    \item[(iv)] and a mixed scenario with branching ratios $(0.3,\, 0.4,\, 0.3)$ for $(e^{\pm}e^{\pm}, \,e^{\pm}\mu^{\pm}, \,\mu^{\pm}\mu^{\pm})$ (\Cref{fig:case_30_40_30}).
\end{itemize}  

The dotted black curves in the figures correspond to our theoretical prediction, while the solid blue curves represent the ATLAS exclusion limits. We find good agreement across all scenarios; for instance, in the purely dielectron case, we obtain a lower mass bound of approximately $m_{H^{\pm\pm}}\simeq 768$~GeV, consistent with the ATLAS result \cite{ATLAS:2017xqs}. The solid yellow curves represents the theoretical prediction considering contributions from $\gamma\gamma$ fusion. Similar agreement is observed for the other decay configurations. 

\begin{figure}[!h]
    \includegraphics[width=0.48\textwidth]{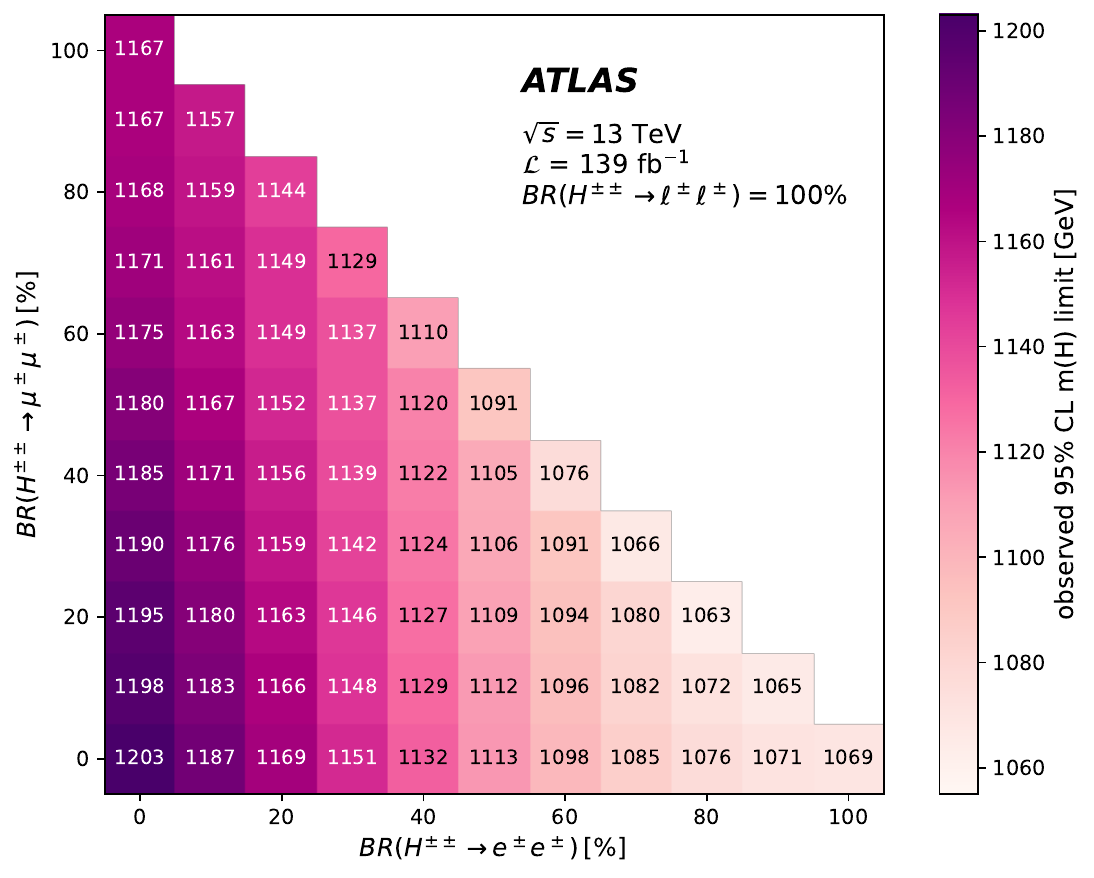}
    \caption{\label{fig:my_label2} 95\% C.L. lower mass bounds on the doubly charged scalar assuming $\text{BR}(H^{\pm\pm}\rightarrow l^{\pm}l^{\pm})=100\%$ for a center-of-mass energy of 13 TeV and integrated luminosity of $139\,\text{fb}^{-1}$. These are estimated values for ATLAS in the next public LHC data release.}
\end{figure}

\paragraph{Projection to the full Run~2 luminosity.}
Assuming that both signal and background event yields scale linearly with luminosity, and that the acceptance and selection efficiencies remain approximately stable, we extrapolate the ATLAS limits to the full Run~2 dataset corresponding to $\mathcal{L}=139~\mathrm{fb}^{-1}$. Under these assumptions, the expected sensitivity scales approximately as $1/\sqrt{\mathcal{L}}$, which we implement by rescaling the published $36.1~\mathrm{fb}^{-1}$ limits. The resulting projected exclusions for the different benchmark branching-ratio scenarios are shown by the dashed curves in \Cref{fig:case_100_0_0,fig:case_0_100_0,fig:case_0_0_100,fig:case_30_40_30} and summarized in \Cref{fig:my_label2}. As in the experimental analyses, decays involving third-generation leptons are not considered.

\paragraph{Comments on Photon--Photon Fusion}

We emphasize that we are considering a model in which the total width of the doubly charged scalar is narrow. We simulated our events for LHC 13 TeV with the inclusion of elastic and inelastic photons using the PDF \textit{LUXlep-NNPDF31\_nlo\_as\_0118\_luxqed} \cite{Manohar:2016nzj,Bertone:2017bme,Manohar:2017eqh} and found no sizable change in the overall production cross section (see Figs. \ref{fig:case_100_0_0}-\ref{fig:case_30_40_30}). In the large-width regime, this contribution was found to be relevant \cite{Crivellin:2018ahj}. This is understood because the large $t$-channel correction to $\gamma\gamma \rightarrow H^{++}H^{--}$ is suppressed.
\section{FCC Reach \label{SecIV_FCC}}

Using our limits for $\sqrt{s}=13$ TeV with $\mathcal{L}=139\,\text{fb}^{-1}$ as a starting point, we assess the FCC-hh reach for benchmarks of $\mathcal{L}=1\,\text{ab}^{-1}$ and $\mathcal{L}=3\,\text{ab}^{-1}$. We do so by assuming that: (i) cross sections (signal and background) scale with $1/M_{H^{\pm\pm}}^2$ and the partonic luminosities; (ii) efficiencies remain roughly constant up to $\sqrt{s}=100$ TeV; (iii) no new background source emerges. Since the signal events from doubly charged scalars feature a distinctive signature, these assumptions are reasonable and have yielded good agreement with data. Using LHC data searches for doubly charged scalars with $\mathcal{L}=4.7\,\text{fb}^{-1}$ and $\sqrt{s}=7$ TeV \cite{ATLAS:2012hi}, one can successfully forecast LHC sensitivity for $\sqrt{s}=13$ TeV and $\mathcal{L}=36.1\,\text{fb}^{-1}$ \cite{ATLAS:2017xqs}. Comparing this forecast with the observed limit, we find good agreement \cite{colliderreach1}.

\begin{figure}[!h]
    \includegraphics[width=0.48\textwidth]{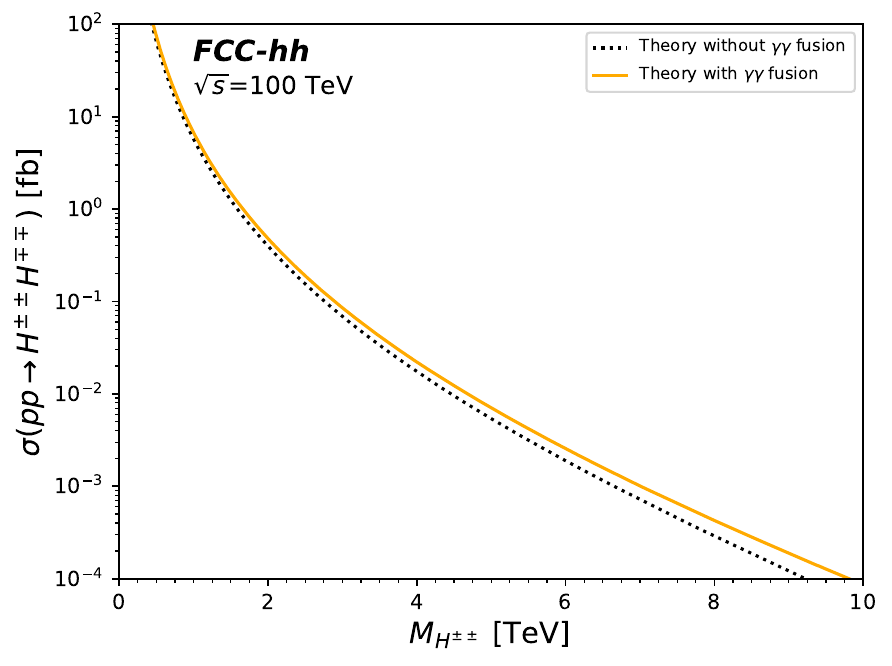}
    \caption{\label{fig:pp100TeV} {\bf} Production cross section $pp\rightarrow H^{\pm\pm}H^{\mp\mp}$ as a function of mass for $\sqrt{s}=100$~TeV.}
\end{figure}

\begin{figure}[!h]
    \includegraphics[width=0.48\textwidth]{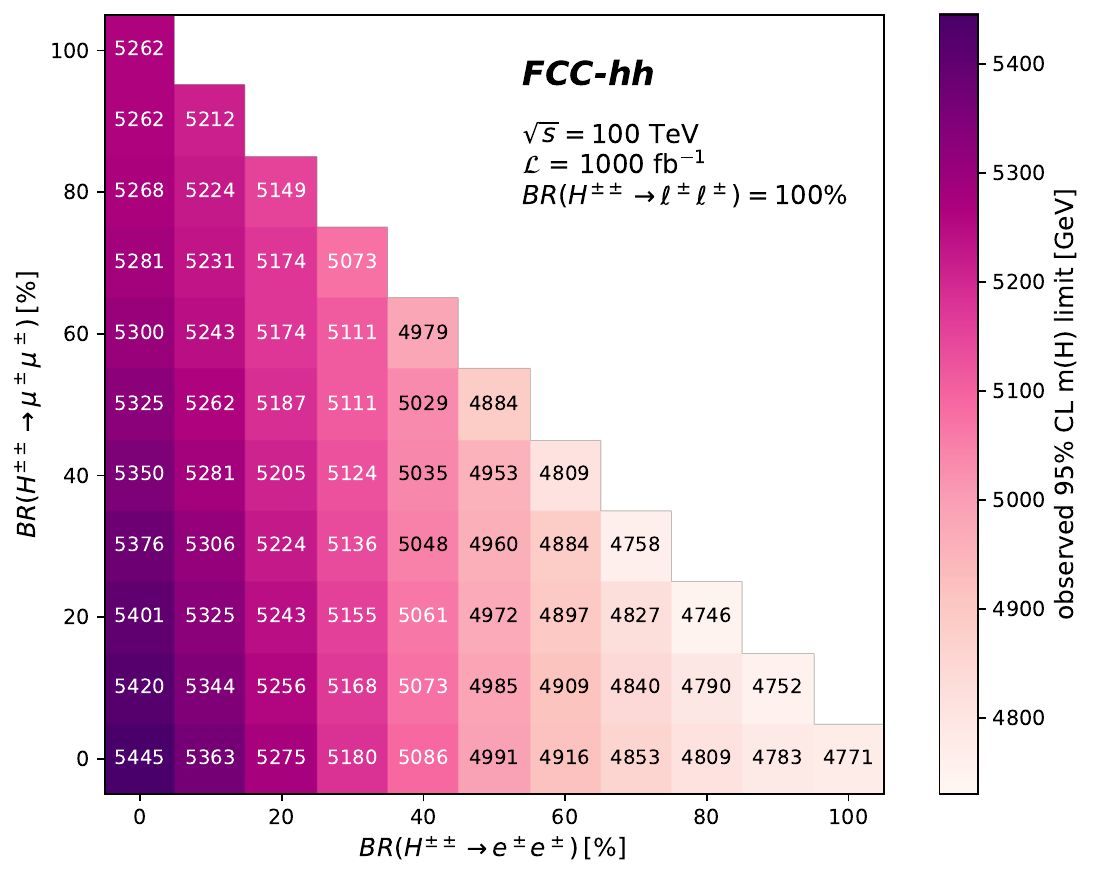}
    \caption{\label{fig:my_label3} Projected lower mass bounds on the doubly charged scalar assuming $\text{BR}(H^{\pm\pm}\rightarrow l^{\pm}l^{\pm})=100\%$ for FCC-hh with $\mathcal{L}=1\,\text{ab}^{-1}$.}
\end{figure}

\begin{figure}[!h]
    \includegraphics[width=0.48\textwidth]{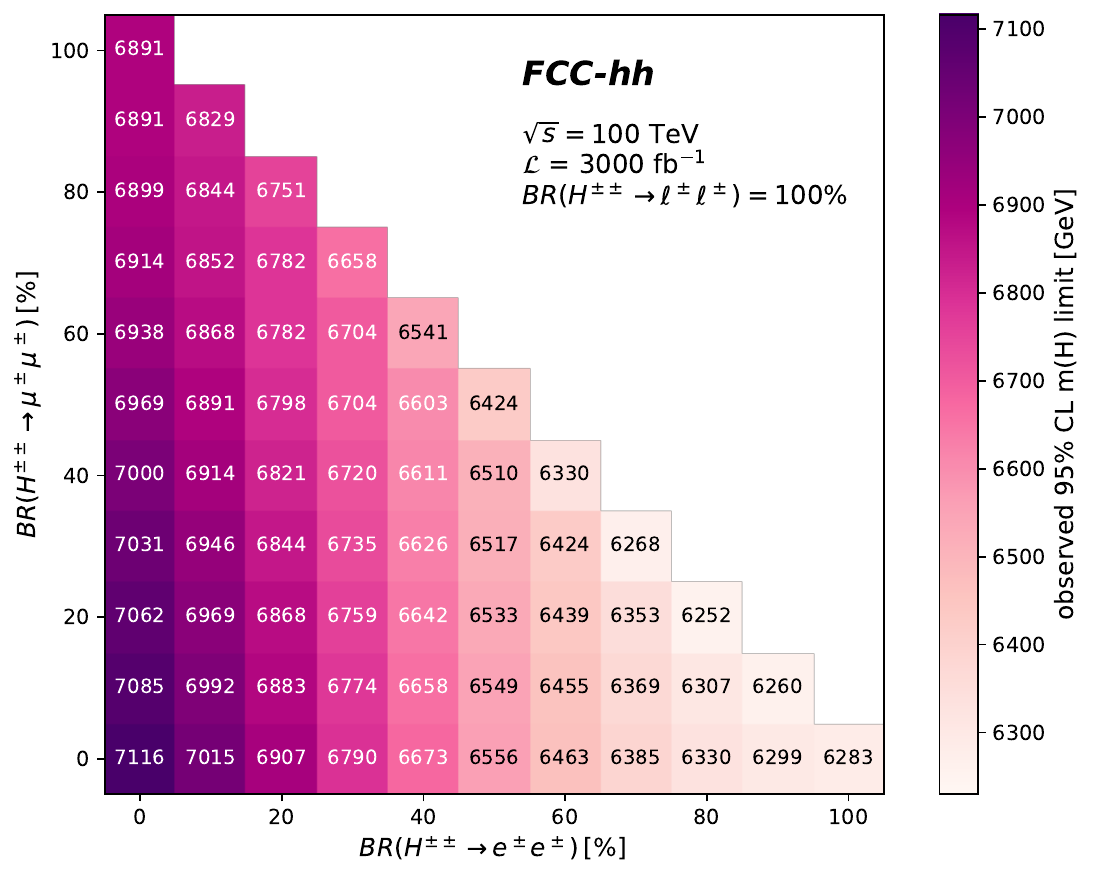}
    \caption{\label{fig:my_label4} Projected lower mass bounds on the doubly charged scalar assuming $\text{BR}(H^{\pm\pm}\rightarrow l^{\pm}l^{\pm})=100\%$ for FCC-hh with $\mathcal{L}=3\,\text{ab}^{-1}$.}
\end{figure}

We estimated the FCC-hh reach for doubly charged scalars for different decay configurations. In \Cref{fig:pp100TeV}, we show the $pp \rightarrow H^{\pm\pm}H^{\mp\mp}$ production cross section at $\sqrt{s}=100$ TeV as a function of the doubly charged scalar mass for $\text{BR}(H^{\pm\pm})=1$. Based on the reasoning described above, we derived the projected constraints in \Cref{fig:my_label3,fig:my_label4} for $\mathcal{L}=1 \ {ab}^{-1}$ and $\mathcal{L}=3 \ {ab}^{-1}$, respectively, covering several decay configurations.

Interestingly, FCC-hh at $\sqrt{s}=100$ and $\mathcal{L}=1 \ {ab}^{-1}$ (\Cref{fig:my_label3}), has the potential to reach masses around 5.4 TeV, which far exceeds the projection of any other proposed future collider \cite{deBlas:2025gyz}. Taking into account on-shell and off-shell production of doubly charged scalars at the International Linear Collider, masses up to 1 TeV can be tested depending on the size of the Yukawa coupling \cite{Nomura:2017abh, Dev:2018upe}. The larger the decay into electrons, the weaker the bound, as a direct consequence of the electron misidentification rate. In particular, we obtain $M_{H^{\pm\pm}}> 4.77$ TeV for $\text{BR}(H^{\pm\pm}\rightarrow e^{\pm}e^{\pm})=100\%$, and $M_{H^{\pm\pm}}> 5.26$ TeV for $\text{BR}(H^{\pm\pm}\rightarrow \mu^{\pm}\mu^{\pm})=100\%$. Thus, the difference in the lower mass limit is indeed sizable.

For completeness, we repeated this procedure for $\mathcal{L}=3\,\text{ab}^{-1}$ and found that in this configuration, masses near 7 TeV can be achieved. Indeed, FCC-hh is undoubtedly the desired machine for new physics discoveries in high-energy physics \cite{Cavaliere:2025ujf}.

\section{Conclusions \label{SecV_Con}}
Motivated by the prevalence of doubly charged scalars in new physics studies connected to neutrino masses, we have assessed the LHC Run II sensitivity with $\mathcal{L}=139\,\text{fb}^{-1}$ to doubly charged scalars decaying into leptons within the narrow-width approximation. In particular, for $\text{BR}(H^{\pm\pm}\rightarrow \mu\mu)=100\%$, we find $M_{H^{\pm\pm}}> 1.17$ TeV (\Cref{fig:my_label2}). Decays into electrons yield weaker limits because of the electron misidentification rate. We subsequently computed the production cross section for $\sqrt{s}=100$ TeV and forecasted the limits for many decay configurations with $\mathcal{L}=1$ and $3\,\text{ab}^{-1}$ (see \Cref{fig:my_label3,fig:my_label4}). For instance, we concluded that FCC-hh can probe masses of approximately 7 TeV for $\mathcal{L}=3\,\text{ab}^{-1}$. It is clear that FCC-hh can probe masses higher than those accessible at other proposed future colliders.

\section*{Acknowledgements}
FSQ thanks the Max Planck Institute for Nuclear Physics and CERN for the hospitality during the final stages of this work. The authors thank Thiago Tomei, Patricia Teles, Sandro Fonseca, Oscar Eboli, Yara Coutinho, and Diego Cogollo for discussions. This work is supported by the Simons Foundation (Award Number: 1023171-RC) and ANID - Millennium Science Initiative Program - ICN2019\_044, and IIF-FINEP grant 213/2024. Y.M. Oviedo-Torres was supported by FONDECYT grant 3250068. FSQ acknowledges FAPESP Grants 2018/25225-9, 2021/01089-1, 2023/01197-4, ICTP-SAIFR FAPESP Grants 2021/14335-0, CNPQ Grants 403521/2024-6, 408295/2021-0, 403521/2024-6, 406919/2025-9, 351851/2025-9. We also thank IIP for the local cluster {\it bulletcluster}, which was instrumental in this work.

\def\bibsection{\section*{References}}
\bibliographystyle{JHEP}
\bibliography{ref}
\end{document}